\documentstyle[aps,graphicx,floats,tighten,preprint]{revtex}

\def\be{\begin{equation}}
\def\ee{\end{equation}}
\def\bea{\begin{eqnarray}}
\def\eea{\end{eqnarray}}

\begin{document}
\title{Electromagnetic Form Factors of the Nucleon in the 
Chiral Soliton Model}
\author{G. Holzwarth\thanks{%
e-mail: holzwarth@physik.uni-siegen.de}}
\address{Fachbereich Physik, Universit\"{a}t Siegen, 
D-57068 Siegen, Germany} 
\maketitle

\begin{abstract}\noindent
Several years ago it was pointed out that the chiral soliton model
allows naturally for satisfactory agreement with the 
experimentally well-determined proton magnetic form factor $G_M^p$.
The corresponding result for the proton electric form factor at that
time was in serious disagreement with the data because the calculated
$G_E^p$ showed as a rather stable feature a zero for $q^2$ near 10
(GeV/c)$^2$ which was hard to avoid for reasonable choices of parameters,
while the data at that time showed no indication for such a behaviour.
Meanwhile, new data have confirmed those $G_E^p$ predictions in a
remarkable way, so it appears worthwhile to have another look at that
model, especially concerning its flexibility with repect to the
electric neutron formfactor $G_E^n$ while trying to maintain the 
satisfactory results for the proton form factors.
\end{abstract} 

\vspace{3cm}
\leftline{PACS numbers:12.39.Fe,13.40.Gp }  

\leftline{Keywords: Nucleon, Electromagnetic form factors, 
Chiral soliton model} 

\newpage

Several years ago it was pointed out \cite{HolZP,Hol} 
that the chiral soliton model allows quite naturally for very
satisfactory agreement with the experimentally well-determined proton
magnetic form factor $G_M^p$ for momentum transfers $q^2$ up to 30
(GeV/c)$^2$. 
The corresponding result for the proton electric form factor at that
time was in serious disagreement with the data because the calculated
$G_E^p$ showed as a rather stable feature a zero for $q^2$ near 10
(GeV/c)$^2$ which was hard to avoid for reasonable choices of parameters,
while the data at that time showed no indication for such a behaviour.
The electric neutron square-radius $\langle r^2 \rangle_E^n$ for the
parametrizations used at that time was too large (typically $\sim$ -0.25
fm$^2$ as compared to the experimental value of -0.114$\pm$0.003
fm$^2$) with a resulting electric neutron form factor rising to a
maximum of about 0.09 as compared to the maximum of the Galster
parametrization of about 0.05. 
Meanwhile, new data have confirmed the predictions for $G_E^p$ in a
remarkable way, so it appears worthwhile to have another look at that
model, especially concerning its flexibility with respect to the
electric neutron formfactor $G_E^n$ while trying to maintain the 
satisfactory results for the proton form factors.

It is well known that the e.m.~form factors obtained from the 
plain standard Skyrme model \cite{Braaten} are insufficient and that
inclusion of vector meson contributions is necessary
\cite{MKW87,Mei93}. There are basically two
simple versions to achieve this: 

Model A \cite{HolZP} : The pionic Skyrme model for the chiral SU(2)-field $U$ 
\be
\label{Skyrme}
{\cal L^{(\pi)}}={\cal L}^{(2)}+{\cal L}^{(4)}
\ee
\be
{\cal L}^{(2)}=\frac{f_\pi^2}{4}\int \left(-trL_\mu L^\mu+m_\pi^2 
tr(U+U^\dagger-2) \right)d^3x,~~~~~
{\cal L}^{(4)}=\frac{1}{32e^2}\int tr[L_\mu,L_\nu]^2 d^3x
\ee
(where $L_\mu$ denotes the chiral gradients $L_\mu 
= U^\dagger \partial_\mu U$ ) is used with its standard constants: 
pion decay constant $f_\pi$=93 MeV, pion mass $m_\pi$=138 MeV, and the
well-established Skyrme parameter $e$=4.25. 
The coupling to the photon field is obtained through the local gauge
transformation $U \to e^{i \epsilon Q}Ue^{-i \epsilon Q}$ with the
charge operator $Q=(\frac{1}{3}+\tau_3)/2$. The isoscalar part of the
coupling arises from gauging the standard Wess-Zumino term.

To incorporate vector meson effects the resulting form 
factors then are multiplied by the factors 
\be
\Lambda_I(q^2)=\lambda_I \left(\frac{m_I^2}{m_I^2+q^2}\right)+(1-\lambda_I)
\ee
with $I$=0,1 for isoscalar and isovector form factors;
$m_0,m_1$ are the $\omega$- and $\rho$-masses $m_\omega=783$ MeV,
$m_\rho=770$ MeV, respectively; so the parameters
$\lambda_0,\lambda_1$ allow for admixing of the vector meson poles to the
purely pionic formfactors. The detailed expressions for the form
factors are given explicitely in~\cite{HolZP}. So, with $e$ kept fixed
at its standard value, this most simple version contains two
parameters: $\lambda_0$ and $\lambda_1$.

Model B \cite{Hol} : The vector mesons are included explicitely as dynamical
degrees of freedom in the lagrangian. In the minimal version the
axial vector mesons are eliminated in chiral invariant way
\cite{Kaymak,SWHH89}. This leaves two gauge coupling constants $g_\rho,
g_\omega$ for $\rho$- and $\omega$-mesons. 
\be
\label{mini}
{\cal L}={\cal L^{(\pi)}}+{\cal L}^{(\rho)}+{\cal L}^{(\omega)}
\ee
\be
{\cal L}^{(\rho)}= \int \left(-\frac{1}{8} tr \rho_{\mu\nu} \rho^{\mu\nu} 
+\frac{m_\rho^2}{4} tr(\rho_\mu
-\frac{i}{2g_\rho}(l_\mu-r_\mu))^2 \right) d^3x, 
\ee
\be
{\cal L}^{(\omega)}=\int \left(-\frac{1}{4} \omega_{\mu\nu} 
\omega^{\mu\nu} +\frac{m_\omega^2}{2} 
\omega_\mu \omega^\mu +3g_\omega \omega_\mu B^\mu \right) d^3x,
\ee
with the topological baryon current $B_\mu=\frac{1}{24 \pi^2}
\epsilon_{\mu\nu\rho\sigma} tr L^\nu L^\rho L^\sigma$, and
$l_\mu=\xi^\dagger \partial_\mu \xi, \;r_\mu=\partial_\mu \xi 
\xi^\dagger$, where $\xi^2=U$. 

The contributions of the vector mesons to the electromagnetic 
currents arise from the local gauge transformations 
\be
\rho^\mu \rightarrow  e^{i \epsilon Q_V}
\rho^\mu e^{-i \epsilon Q_V}  +\frac{Q_V}{g_\rho} 
\partial^\mu \epsilon,~~~~~~~
\omega^\mu \rightarrow  \omega^\mu 
+\frac{Q_0}{g_0} \partial^\mu \epsilon
\ee
(with $Q_0=1/6 \ , \ Q_V=\tau_3 /2$). 
The resulting form factors are expressed in terms of three static and 
three induced profile functions which characterize the rotating 
hedgehog soliton with baryon number $B=1$.

Because the Skyrme term ${\cal L}^{(4)}$ at least partly accounts for
static $\rho$-meson effects its strength in Model B must be strongly
reduced, or could even be omitted. 
So, in this model we consider $e$ as an additional parameter.
The coupling constant $g_\rho$ can be fixed by the KSRF
relation $g_\rho=m_\rho/(2\sqrt{2} f_\pi)$ = 2.925, but small
deviations from this value are tolerable.
The $\omega$-mesons introduce two gauge coupling constants, $g_\omega$
to the baryon current in ${\cal L}^{(\rho)}$, and $g_0$ for the
isoscalar part of the charge operator. 
Within the $SU(2)$ scheme we can
in principle allow $g_0$ to differ from $g_\omega$ and thus 
exploit the freedom in the e.m. coupling of the isoscalar
$\omega$-mesons. However, as the isoscalar part of the
electromagnetic current is given by the baryonic current, it is
natural to expect $g_\omega \approx g_0$. 

A difficulty of all nucleon models is to relate the form
factors evaluated in the nucleon rest frame to their momentum-transfer
dependence in the Breit frame moving relative
to the rest frame with velocity $v$, with
\be
\gamma^2=(1-v^2)^{-1} = 1 + \frac{q^2}{(2M)^2},
\ee
where $M$ is the nucleon mass.
Unfortunately, the simple boost prescription ~\cite{Licht,Ji91}
\be
\label{boost}
G_M^{Breit} (q^2) = \gamma^{-2}\; G^{rest}_M (
\gamma^{-2}\;q^2),~~~~~~~~~ 
G_E^{Breit} (q^2) = G^{rest}_E ( \gamma^{-2}\; q^2)  
\ee 
has a serious flaw: it generally violates the superconverge law
expected for nucleon formfactors \cite{Matveev}
\be
q^2 G^{Breit} (q^2) \to 0~~~~~\mbox{ for}~~~q^2\to \infty.
\ee
This is due to the fact that the boost in Eq.(\ref{boost})
maps $G^{rest}(4M^2)~\to~G^{Breit}(q^2\to\infty)$,
and $G^{rest}(4M^2)$, although being very small, generally
does not vanish exactly. This shows up, of course, in a very drastic way,
if the resulting formfactors are divided by the standard dipole
\be
G_D(q^2)=1/(1+q^2/0.71)^2,
\ee 
which is the common way to present them. 
So it is vital for a comparison with experimentally determined
form factors for $q^2 \gg M^2$ to modify the boost prescription
in such a way that agreement with the data for $G_M^p/(\mu_p G_D)$ at the
highest available values of $q^2$ is improved.
A simple way to achieve this is to allow the
mass $M$ in the boost Eq.(\ref{boost}) to be larger than the
experimental value of the nucleon mass, which further
decreases the absolute value of $G^{rest}_M(4M^2)$. 
One might even argue this to be
consistent with the soliton model because the classical soliton masses 
typically are around 1.5 GeV.
Further improvement of the high-$q^2$ behaviour could be achieved
by enforcing superconvergence through subtraction of the small
constants $G^{rest}(4 M^2)$ from the form factors
as described in~\cite{HolZP}. In any case, however, the high-$q^2$
behaviour is not a profound consequence of the model but rather 
reflects the boost prescription. So, in the following, we do not
enforce superconvergence.

Of course, it is unfortunate that in this way the physical
information provided by the high-$q^2$ limit of $G_M^p/(\mu_p G_D)$ is lost,
but there is no hope anyway, why such low-energy effective models
should give a profound answer for that high-$q^2$ limit.
However, the functional form of the different form factors relative to each
other remains largely intact. Therefore in both versions, Model A and 
Model B, we introduce the boost mass $M$ in Eq.(\ref{boost}) as one
additional parameter to adjust to the high-$q^2$ data for
$G_M^p/(\mu_p G_D)$.

\begin{figure}[h]
\begin{center}
\includegraphics[width =7cm,height=8cm,angle=-90]{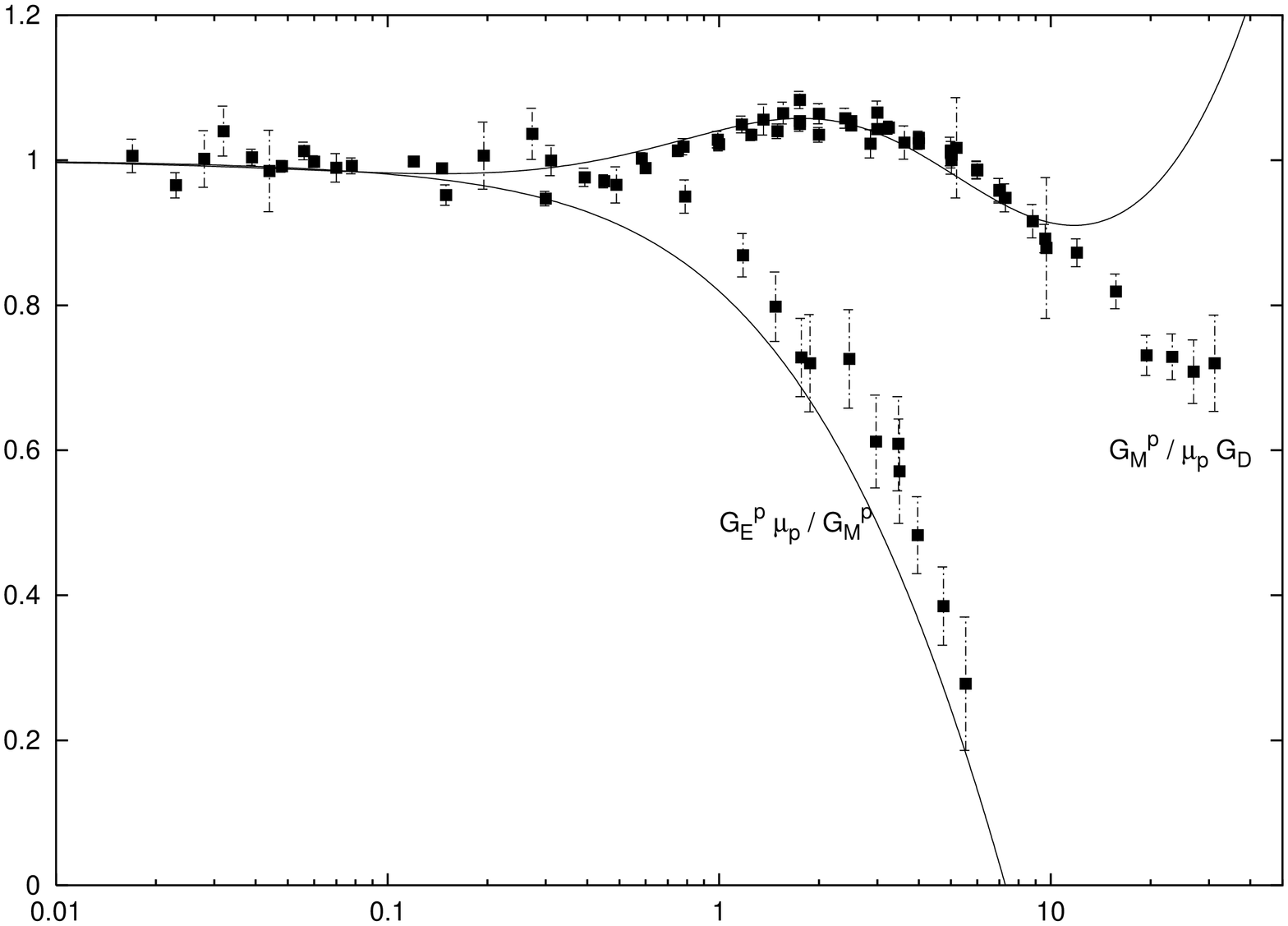}
\includegraphics[width =7cm,height=8cm,angle=-90]{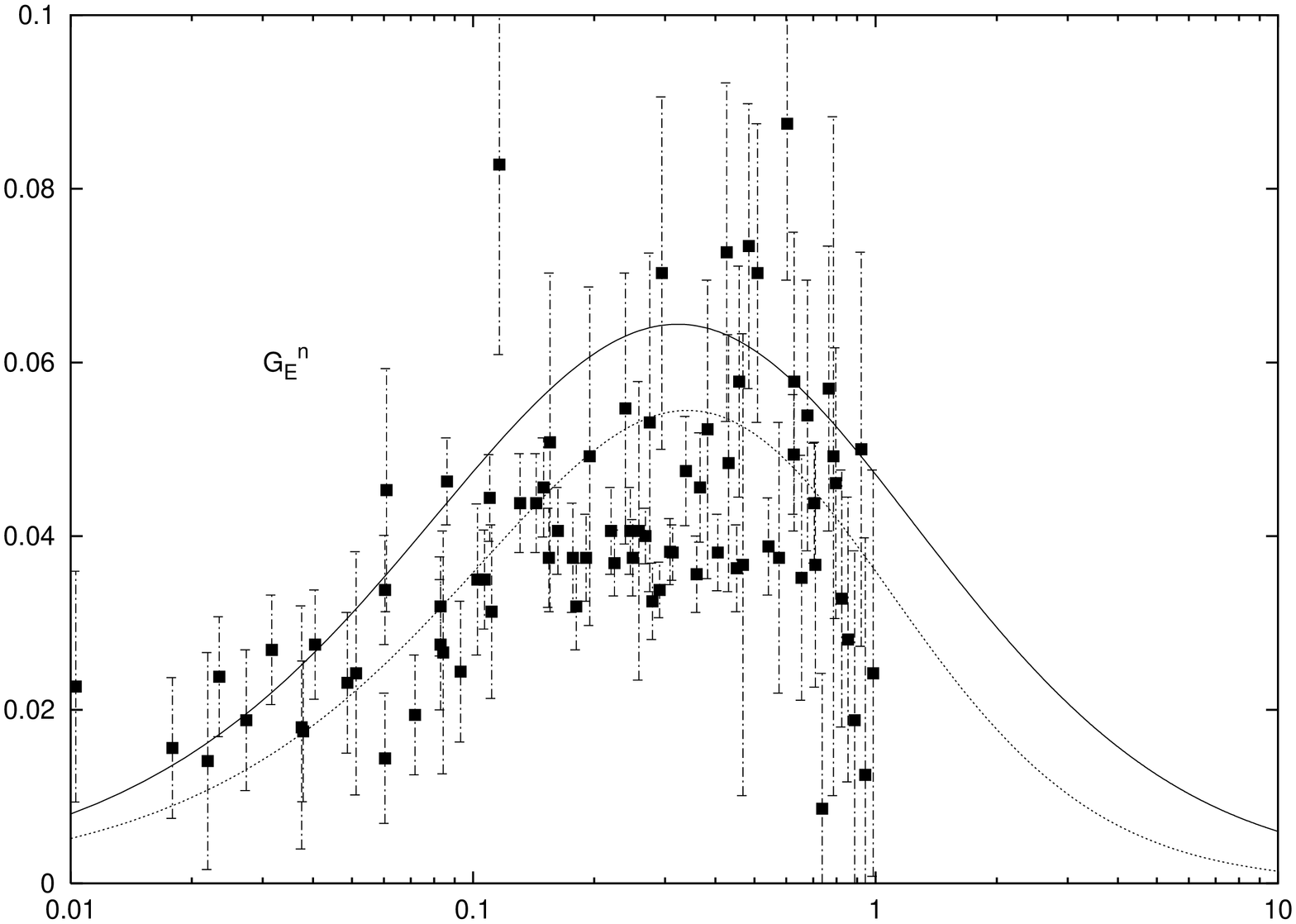}
\end{center}
\caption[]{Magnetic and electric formfactors of the proton 
$G_M^p/(\mu_p G_D)$ and $G_E^p\mu_p /G_M^p$ (left)
and electric formfactor $G_E^n$ of the neutron (right),
for Model A with $\lambda_0$=0.92, $\lambda_1$=0.78, and $M$=1.5 GeV.
The data for $G^p$ are from 
\cite{Jones,Gayou,Hoehler,Sill,Andivahis,Walker}. 
The data for $G_E^n$ are from \cite{Platchkov} for
Paris potential and from \cite{Galster} for Lomon wavefunction, the 
dotted line is the corresponding Galster parametrization. }
\label{ModelA}
\end{figure}

The parameters $\lambda_0$ and $\lambda_1$ in Model A determine
the amount of vector dominance in the isoscalar and iosvector channels,
respectively. Their difference, $\lambda_0-\lambda_1$, therefore 
is crucial for the magnitude of the electric neutron formfactor
$G_E^n$. In~\cite{HolZP} they were taken equal for simplicity
($\lambda_0=\lambda_1=0.75$), which resulted in the quadratic neutron
radius being much too large. Improving  on this
point requires stronger vector dominance for the
$\omega$-meson ($\lambda_0 \to 1)$ with $\lambda_1$ still around 0.75.
We present in Fig.~\ref{ModelA} results for Model A with $\lambda_0=0.92, 
\lambda_1=0.78$, and $M$=1.5 GeV. The form of $G_E^n$
is very similar to the Galster parametrization \cite{Galster}, 
but still exceeds it 
by about 20\% near the maximum. Attempts to further lower it 
(by increasing the difference $\lambda_0-\lambda_1$) result in
simultaneous decrease of $G_E^p \mu_p /G_M^p$, moving it further to the
left of the recent data set of \cite{Jones,Gayou}.
So, in this very restricted model
we find it difficult to bring $G_E^n$ down to the Galster result,
while maintaining overall agreement with both proton form factors.
(The sharp rise of $G_M^p/(\mu_p G_D)$ beyond 10 (GeV/c)$^2$ is due
to the small finite value of $G^{rest}(4 M^2)$. It could be removed 
by enforcing superconvergence.)

\begin{figure}[h]
\begin{center}
\includegraphics[width =7cm,height=8cm,angle=-90]{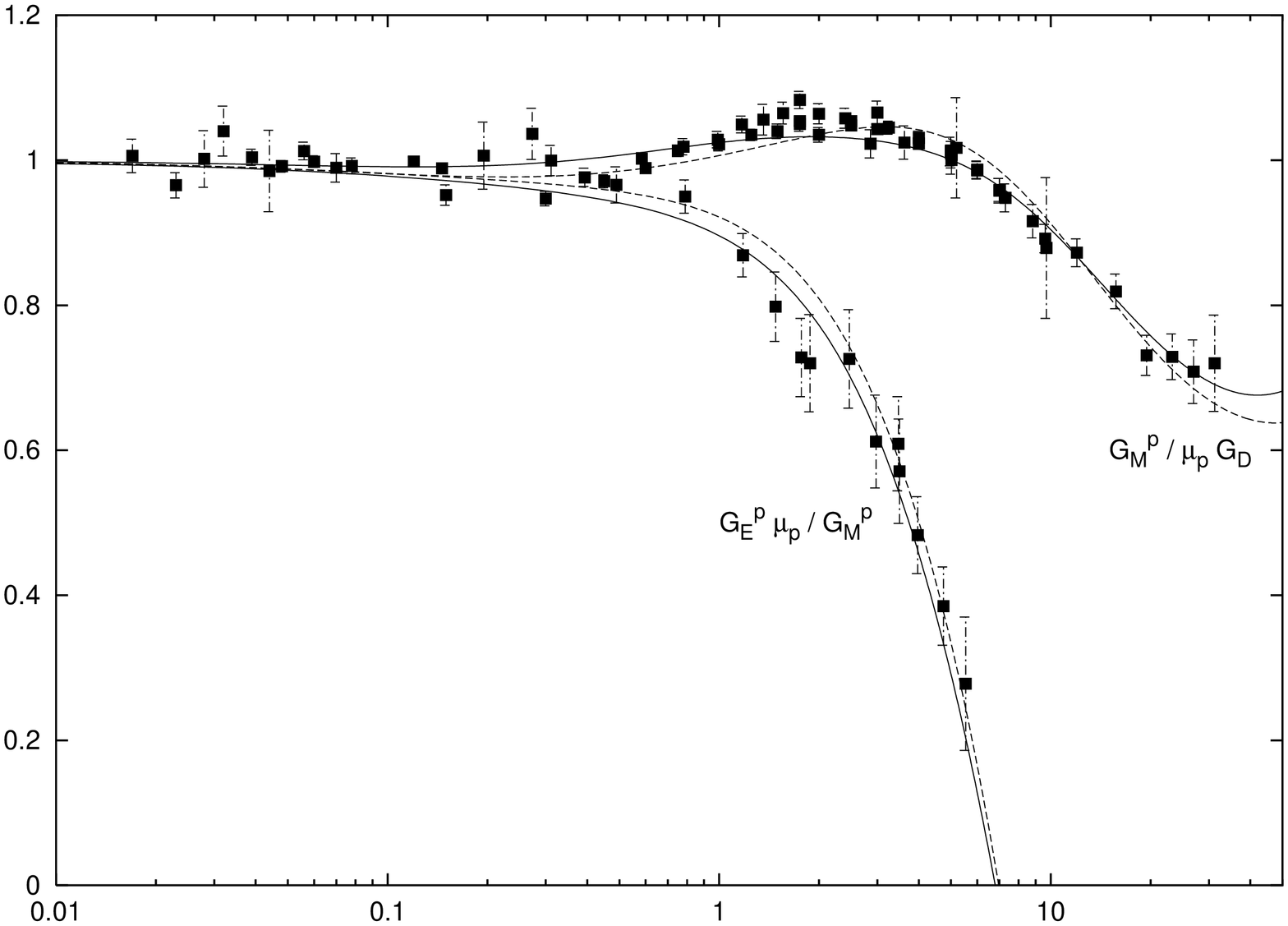}
\includegraphics[width =7cm,height=8cm,angle=-90]{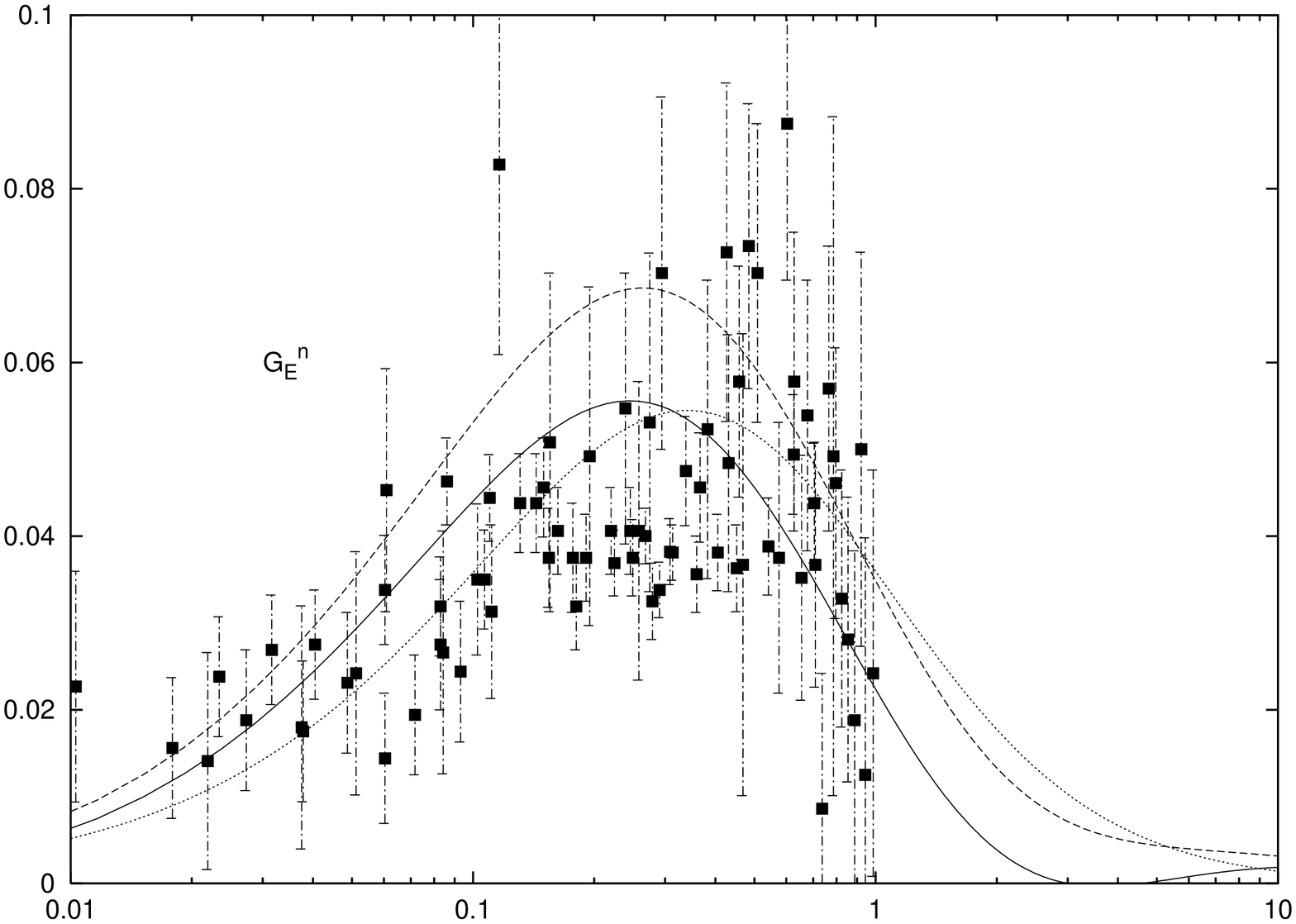}
\end{center}
\caption[]{The same as Fig.\ref{ModelA} for Model B:
 Fit B1 (dashed lines): $e$=12, $g_\rho$=2.6, $g_\omega$=$g_0$=1.4, and
$M$=1.89 GeV; Fit B2 (full lines): $e$=12, $g_\rho$=2.64, $g_\omega$=0.9,
$g_0$=1.1$g_\omega$, and $M$=2.1 GeV. The dotted line for $G_E^n$ again
is the Galster parametrization. }
\label{ModelB}
\end{figure}

In Model B the amount of vector dominance for
the $\rho$-meson is fixed by the ratio $g_\rho/(m_\rho f_\pi)$, 
while for the $\omega$-mesons it is determined by 
$g_\omega/g_0=\lambda_0$. So, the results from Model A imply
that the constraint $g_\omega=g_0$ in Model B should lead to
better results for $G_E^n$, which indeed proves
to be the case. For satisfactory overall agreement of the proton form
factors we find it helpful to keep a small Skyrme term in the 
lagrangian (note that it is $\sim$$e^{-2}$, so with $e$=12
it has about 10\% of its standard strength). We present in
Fig.~\ref{ModelB} two fits: (B1) with $e$=12, $g_\rho$=2.6, 
$g_\omega$=$g_0$=1.4, and $M$=1.89 GeV; and (B2): with $e$=12,
$g_\rho$=2.64, $g_\omega$=0.9, $g_0$=1.1$g_\omega$, and $M$=2.1 GeV.
Again it proves difficult to further lower the electric neutron
form factor, with the ratio $g_\omega/g_0$=1 fixed (in fit B1), 
and trying to keep $G_E^p \mu_p /G_M^p$ within the data. The form of
$G_E^n$ in this 
case differs slightly from the Galster form, the maximum is shifted
to lower $q^2$, and the following decrease is steeper, so $G_E^n$
for $q^2 > 1$(GeV/c)$^2$  is smaller than the Galster result. 
However, $G_E^n$ is very sensitive to the ratio $g_\omega/g_0$ and 
allowing for a 10\% increase (fit B2) brings its maximum 
down to the Galster value. Readjustment of $g_\rho$ and $g_\omega$
allows to maintain the agreement with both proton form factors.
The resulting $G_E^n$ develops a small dip beyond its main decrease
so it actually passes through zero near $q^2 \approx 3$(GeV/c)$^2$.
For small $q^2$, $G_E^n$ still exceeds the Galster parametrization
(because the maximum is shifted to the left), so the 
absolute values of the resulting neutron square radii are still too
large (cf. Table 1).

\begin{figure}[h]
\begin{center}
\includegraphics[width =8cm,height=11cm,angle=-90]{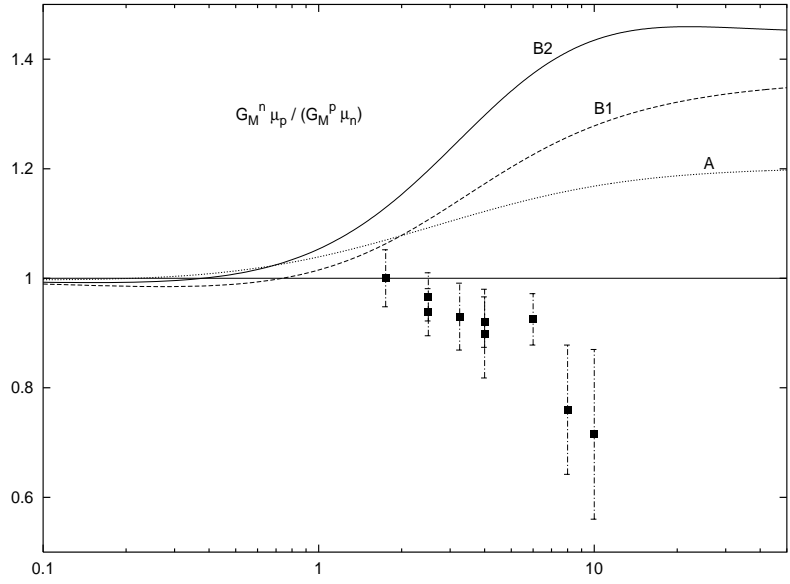}
\end{center}
\caption[]{The ratio of normalized magnetic neutron and proton 
form factors $G_M^n \mu_p /(G_M^p \mu_n)$ for model A (dotted line)
and for model B (fit B1: dashed line, fit B2: full line).
The data are from \cite{Rock,Lung}.}
\label{gmnAB}
\end{figure}

It is of interest to also look at the magnetic neutron form factor
$G_M^n$. In order to get rid of the problems with superconvergence
we consider the ratio of the normalized proton and neutron form
factors $G_M^n \mu_p /(G_M^p \mu_n)$. In Fig.\ref{gmnAB} we present
these ratios for both models, together with data from \cite{Rock,Lung}.
Both models consistently predict this ratio to increase above 1 
by 20-40\% for $q^2 > 1$(GeV/c)$^2$ . This increase is the more pronounced
the lower the value of $G_E^n$ near 1 (GeV/c)$^2$ is. The present
data do not show such an increase, in fact they
indicate the opposite tendency. This conflict was already noticed in
\cite{HolZP}. 

In Table 1 we list the quadratic radii and magnetic moments as they
arise from the three fits given above. Notoriously low are the
magnetic moments, as is well known in chiral soliton models.
Quantum corrections may partly be helpful in this respect (see
\cite{MeiWall97}), as they certainly are for the absolute values of the
masses . 
Of course both models can be extended; the addition of 6th order terms
in Model A, the explicite inclusion of axial vector mesons in Model B 
provide more flexibility through additional parameters. It is,
however, remarkable that in their minimal versions as described above 
they are able to provide quite satisfactory results for both
proton and the electric neutron form factors. In fact, the sharp drop
in $G_E^p$ was predicted by these models, and it would be very
interesting to have also new data for $G_M^n$ concerning the conflict
indicated in Fig.\ref{gmnAB}. Evidently, the weakest point of these
considerations is the transition from the rest- to the Breit frame.
Although it looks quite natural, the Ji-prescription Eq.(\ref{boost})
is very unsatisfactory, and it would be highly desirable to have
superconvergence incorporated in a cogent way. As long as this
problem has not been settled there is little hope to gain profound
insight from high-$q^2$ ( $>$10(GeV/c)$^2$ ) data for e.m.form factors.

\begin{table}[h]
\begin{tabular}{|c|c|c|c|c|} 
& & & &\\
~~~~~~~~~&~~Model A~~~&~~Model B1~~&~~Model B2~~&~~~~~Exp.~~~~~\\[.2cm] \hline
& & & &\\
~~~$\langle r^2 \rangle ^p_E$~~~  & 0.795 & 0.807 & 0.782 & 0.74 \\[.2cm]
~~~$\langle r^2 \rangle ^p_M$~~~  & 0.713 & 0.738 & 0.708 & 0.74 \\[.2cm]
~~~$\langle r^2 \rangle ^n_E$~~~  & -0.200~ & -0.238~ & -0.203 & -0.114
\\ [.2cm]
~~~$\langle r^2 \rangle ^n_M$~~~  & 0.729  & 0.776  &  0.739  & 0.77~ \\[.2cm]
~~~$\mu_p$~~~       & 1.78~ & 1.71~ & 1.49 & 2.79~ \\[.2cm]
~~~$\mu_n$~~~       & -1.42~~ & -1.27~~ & -1.05~ & -1.91~~ \\[.3cm] 
\end{tabular}\hspace*{\fill}\\[.3cm]
\caption[]{Nucleon quadratic radii and magnetic moments as obtained
from Models A and B, for the fits given in the text, 
compared to their experimental values \cite{Simon}. }
\end{table}


\end{document}